\documentclass[a4paper,twocolumn,pra,showpacs]{revtex4}
\pdfoutput=1

\usepackage{amsmath,amsfonts,amssymb}
\usepackage{graphicx}
\usepackage{enumerate}
\usepackage{bbm}

\newcommand{\eg}{\emph{e.g.}}

\newcommand{\se}[1]{\ensuremath{^{\text{#1}}}}
\newcommand{\si}[1]{\ensuremath{_{\text{#1}}}}
\newcommand{\vect}[1]{\ensuremath{\vec{#1}}}

\newcommand{\1}{\ensuremath{\mathbbm{1}}}

\newcommand{\twoFone}{\ensuremath{{_2}F_1}}

\begin{document}
 
\title{Electrostatics of Gapped and Finite Surface Electrodes}
\date{\today}
\author{Roman Schmied}
\email{Roman.Schmied@mpq.mpg.de}
\affiliation{Max Planck Institute of Quantum Optics, Garching, Germany}


\begin{abstract}
	We present approximate methods for calculating the three-dimensional electric potentials of finite surface electrodes including gaps between electrodes, and estimate the effects of finite electrode thickness and an underlying dielectric substrate. As an example we optimize a radio-frequency surface-electrode ring ion trap, and find that each of these factors reduces the trapping secular frequencies by less than 5\% in realistic situations. This small magnitude validates the usual assumption of neglecting the influences of gaps between electrodes and finite electrode extent.
\end{abstract}
 
\maketitle

Trapped ions have shown to be excellent candidates for building quantum information processing experiments~\cite{Cirac1995,SchmidtKaler2003,Blatt2008,Buluta2009}. It is possible to trap individual ions~\cite{Dehmelt1990,Leibfried2003}, manipulate their motional and internal degrees of freedom with outstanding fidelity~\cite{Leibfried2003}, and read out their full quantum-mechanical state~\cite{Haeffner2005}. In order to construct small and scalable setups of trapped ions it is necessary to build and improve radio-frequency (rf) Paul microtraps constructed from surface electrodes~\cite{Seidelin2006,Amini2008}. The design of surface electrodes is well understood~\cite{Wesenberg2008,House2008,Schmied2009} as long as several simplifying assumptions are made. In this work we discuss two of these assumptions, present techniques for more accurate calculations leading to better designs, and give estimates of how good these assumptions are in practice.

The particular assumptions under study in this work are (i) that there are no finite gaps between surface electrodes and (ii) that the surface electrodes extend to infinity and cover the entire plane. Such planar infinite gapless electrodes are assumed in order to simplify theoretical work~\cite{Wesenberg2008,House2008,Schmied2009} and to achieve analytic and general results. By contrast, experimental setups are often simulated with boundary-element (BEM) or similar numerical methods~\cite{Wrobel,Liu2006}, which are much more demanding and only give narrow results for specific setups. All of the results achieved in this work can be further improved upon by such numerical methods; our goal here is merely to estimate the size of the corrections due to the aforementioned assumptions and to suggest fast approximations for taking the leading corrections into account during the design phase of an ion trap setup.

This article is organized as follows. In Section~\ref{sec:2D3D} we summarize the tools used for calculating the electrostatics of planar electrodes. Section~\ref{sec:gaps} looks at the influence of gaps between electrodes, and Section~\ref{sec:infinite} at the influence of the finiteness of the surrounding electrode.

\section{surface charge density}
\label{sec:2D3D}

The mathematical tools for calculating the electric fields of planar infinite gapless electrodes have been presented in detail elsewhere~\cite{Oliveira2001,Wesenberg2008}. They assume that there is an infinite plane in space, here the $x y$ plane, on which the electric potential $\Phi(\vect{r})$ is fully known since it is determined from the shapes of the gapless surface electrodes which are individually connected to externally generated voltage sources (dc or rf). This potential is then mathematically extended away from the electrode plane through Laplace's equation $\nabla^2\Phi(\vect{r}) = 0$ satisfied in vacuum. In all generality this extension can be expressed as a surface integral
\begin{equation}
	\label{eq:Phi23}
	\Phi(x,y,z) = \int_{\mathbb{R}^2} \phi(x',y') G(x-x',y-y',z) \text{d}x' \text{d}y',
\end{equation}
with the surface potential $\phi(x,y)\equiv\Phi(x,y,0)$ and the Green's function
\begin{equation}
	\label{eq:Green}
	G(x,y,z) = \frac{|z|}{2\pi(x^2+y^2+z^2)^{3/2}}.
\end{equation}
Equation~\eqref{eq:Phi23} is sufficient for performing calculations under the approximations of infinite electrodes and no gaps, since the integral is completely determined through the full knowledge of $\phi(x,y)$ and thus the potential $\Phi(\vect{r})$ can be calculated anywhere in space. Many other tools developed for such planar infinite gapless electrodes are derivations of Eq.~\eqref{eq:Phi23}.

Both approximations studied here, gaps between electrodes and empty space surrounding a finite electrode, have in common that their relaxation leads to areas in the $x y$ plane where the electric potential $\phi(x,y)$ is not known \emph{a priori} and hence Eq.~\eqref{eq:Phi23} can no longer be used without first determining the surface potential in the open areas. Here we derive additional constraints on the potential $\Phi(\vect{r})$ which are still satisfied even if such open (electrode-free) areas are present in the $x y$ plane.

If we view the planar electrodes as an infinitely thin sheet floating in vacuum at $z=0$, Eq.~\eqref{eq:Phi23} yields a symmetric potential $\Phi(x,y,-z) = \Phi(x,y,z)$. We define the surface charge density of the sheet~\cite{Jackson}
\begin{equation}
	\label{eq:sigma}
	\sigma(x,y) = -2\varepsilon_0Ê\lim_{z\to 0^+} \frac{\partial \Phi(x,y,z)}{\partial z},
\end{equation}
where $\varepsilon_0$ is the permittivity of free space. As long as the assumptions of no gaps and infinite electrodes are satisfied, $\sigma(x,y)$ can take arbitrary values on the electrodes since charge carriers in metallic electrodes will move to wherever necessary for generating the surface potential $\phi(x,y)$. However, whenever there is a hole in the surface electrodes, for instance within a gap between two surface electrodes, $\sigma(x,y)$ must be zero in this hole since charge carriers cannot exist in the vacuum outside of metallic electrodes. This condition gives us a way of calculating the three-dimensional electric potential of gapped and finite planar electrodes. It is important to note that the surface charge density~\eqref{eq:sigma} and the condition that it vanish outside of the physical electrodes are only valid when the electrodes are infinitely thin and the potential symmetry $\Phi(x,y,-z) = \Phi(x,y,z)$ is satisfied. In Section~\ref{sec:thickness} we study the effect of thick electrodes as well as substrate dielectrics on the conclusions reached from thin-electrode calculations.

The surface charge density (constrained to zero outside of the surface electrodes) and the surface electric potential (constrained to given values inside of the surface electrodes) are related through
\begin{subequations}
\begin{align}
	\label{eq:sigma2phi}
	\phi(x,y) & = \phi_0 + \frac{1}{4\pi\varepsilon_0} \int_{\mathbb{R}^2}
		\frac{\sigma(x',y') \text{d}x' \text{d}y'}{\sqrt{(x-x')^2+(y-y')^2}},\\
	\label{eq:phi2sigma}
	\sigma(x,y) & = -\frac{\varepsilon_0}{\pi}Ê\lim_{z\to 0} \int_{\mathbb{R}^2}
		\phi(x',y') \text{d}x' \text{d}y'\nonumber\\
		& \times \frac{(x-x')^2+(y-y')^2-2z^2}{[(x-x')^2+(y-y')^2+z^2]^{5/2}}.
\end{align}
\end{subequations}
The order of the limit and integral in Eq.~\eqref{eq:phi2sigma} can in general not be interchanged. In order to compute the surface charge density from the surface potential, it is necessary to compute a related quantity (the $z$ component of the electric field) for $z\neq0$ and \emph{then} take the limit $z\to 0$. This makes many calculations involving gapped or finite electrodes much more demanding than they first appear. In practice an approximation is often sufficient: defining a small region $\Delta(x,y)$ in the $x y$ plane close to the point $(x,y)$, we can split the integral into two parts, integrating over $\Delta(x,y)$ and $\mathbb{R}^2\setminus\Delta(x,y)$ separately. In the latter part the limit and integral may be interchanged, thus eliminating the need for a three-dimensional intermediate; in the former part a series expansion of $\phi(x',y')$ around the point $(x,y)$ is usually sufficient. If $\Delta(x,y)$ is a square of sides $2d\times2d$ centered at $(x,y)$, then the local (singular) contribution to Eq.~\eqref{eq:phi2sigma} is found from
\begin{multline}
	\label{eq:phi2sigma_approx}
	\lim_{z\to 0} \int_{x-d}^{x+d} \text{d}x' \int_{y-d}^{y+d} \text{d}y'
	\phi(x',y')\\
	\times \frac{(x-x')^2+(y-y')^2-2z^2}{[(x-x')^2+(y-y')^2+z^2]^{5/2}}\\
	= -\frac{4\sqrt{2}}{d} \phi(x,y)
	+ 2d\sinh^{-1}(1) \nabla^2\phi(x,y)
	+ \mathcal{O}(d^3).
\end{multline}
As $d\to0$ the first term of the series in Eq.~\eqref{eq:phi2sigma_approx} diverges but is cancelled out by an equal and opposite divergent term in the integral over $\mathbb{R}^2\setminus\Delta(x,y)$.

\section{gaps between electrodes}
\label{sec:gaps}

Analytic studies of ion traps built with surface electrodes are often simplified to gapless electrodes~\cite{Wesenberg2008,House2008,Schmied2009}. As a first improvement over this approximation Ref.~\cite{House2008} introduced an interpolation method for calculating the influence of gaps between electrodes to lowest order. We show here that this method fails to take into account the dominant ``gap polarization'' induced by surrounding electrodes, and propose an improved method.

\subsection{infinite straight gap}
\label{sec:infinitegap}

\begin{figure}
	\includegraphics[width=8.5cm]{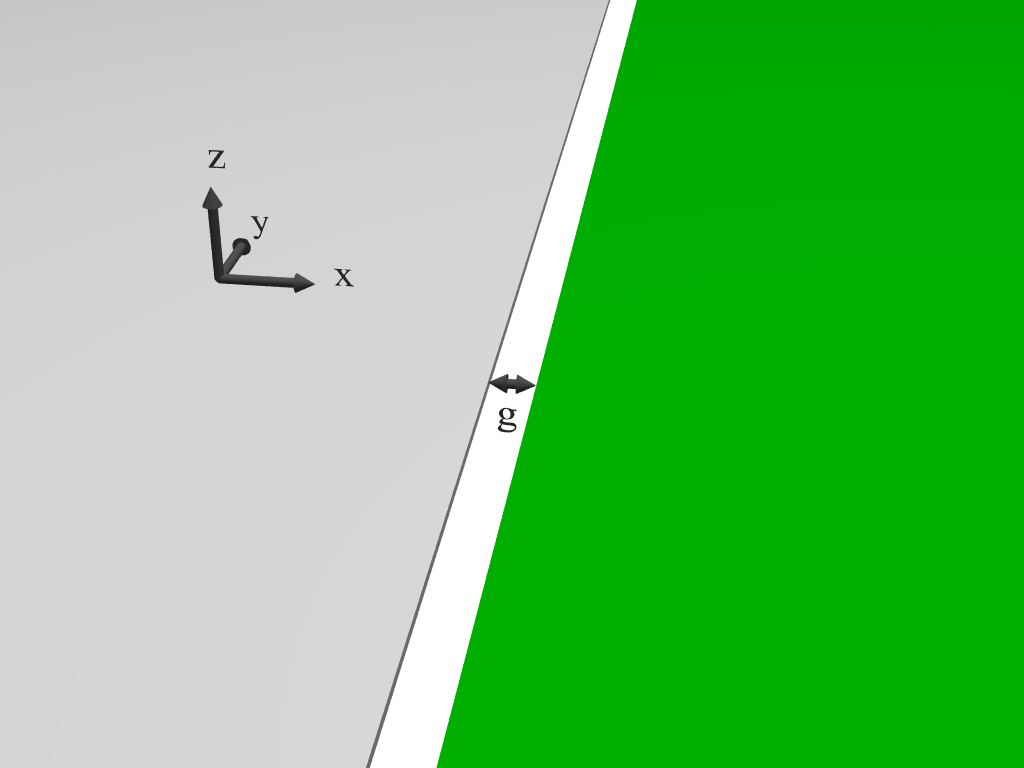}
	\caption{Perspective view of an infinitely long straight gap of width $g$ between two semi-infinite planar electrodes. The left (gray) electrode at $x<-g/2$ is grounded; the right (green) electrode at $x>g/2$ is held at unit potential.}
	\label{fig:infgap}
\end{figure}

\begin{figure}
	\includegraphics[width=8.5cm]{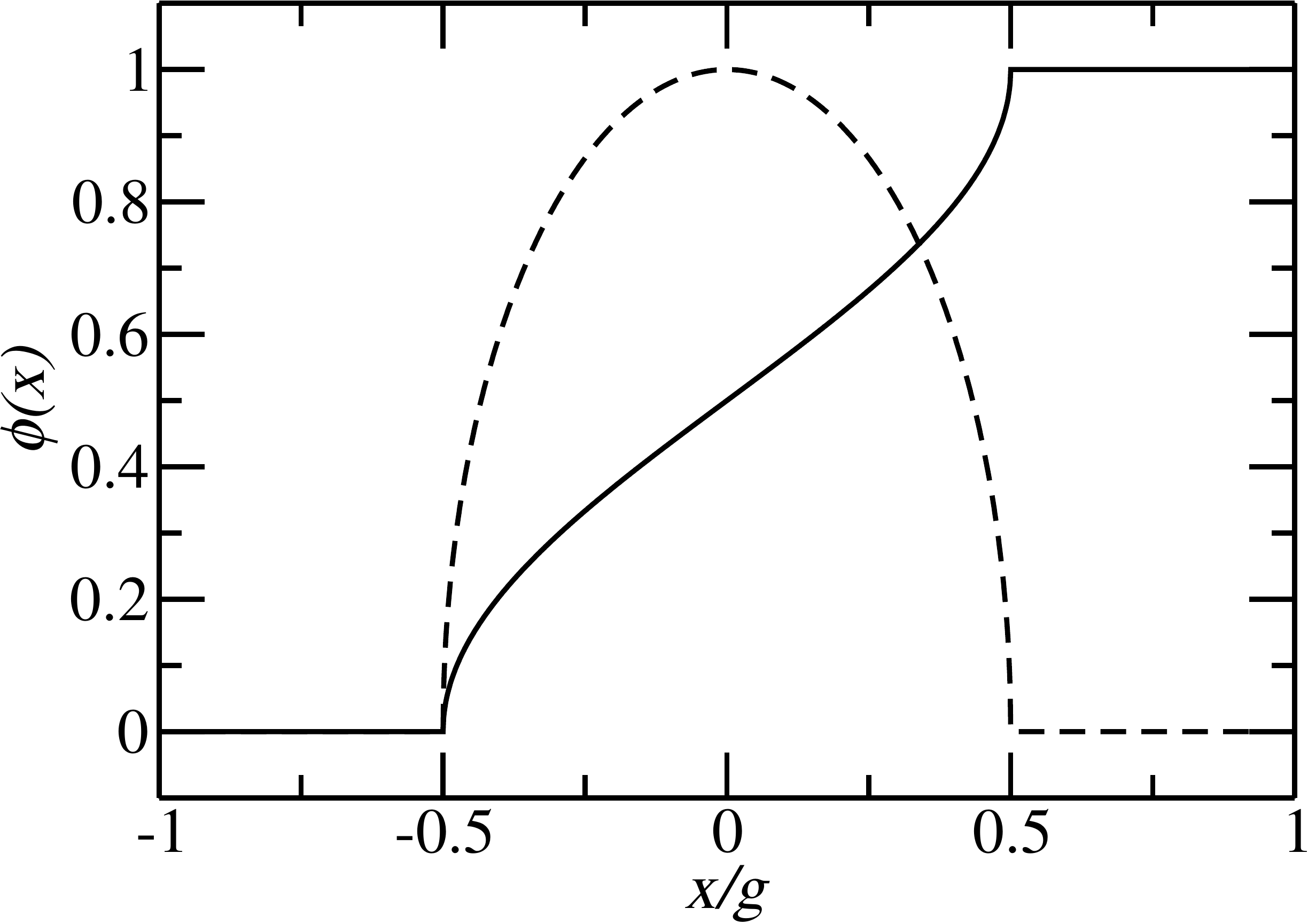}
	\caption{Scaled surface electric potentials in the setup of Fig.~\ref{fig:infgap}. Solid line: gap interpolation potential for thin electrodes, Eq.~\eqref{eq:gappot}. Dashed line: gap polarization potential, Eq.~\eqref{eq:ellpot}.}
	\label{fig:gappot}
\end{figure}

We begin our discussion of gaps between surface electrodes with the simplest possible case: an infinite straight gap between two thin infinite semi-planes at different potentials (Fig.~\ref{fig:infgap}). Its scaled surface potential
\begin{equation}
	\label{eq:gappot}
	\phi_g\se{gap}(x) = \begin{cases}
		0 & \text{if $x\le-g/2$}\\
		\frac12 + \frac{1}{\pi} \sin^{-1}\frac{x}{g/2} & \text{if $|x|<g/2$}\\
		1 & \text{if $x\ge g/2$}
	\end{cases}
\end{equation}
is plotted in Fig.~\ref{fig:gappot}. Its associated surface charge density is calculated from Eq.~\eqref{eq:phi2sigma},
\begin{equation}
	\label{eq:gapscd}
	\sigma_g\se{gap}(x) = \begin{cases}
		-\frac{4 \varepsilon_0}{\pi\sqrt{4x^2-g^2}} & \text{if $x<-g/2$}\\
		0 & \text{if $|x|<g/2$}\\
		\frac{4 \varepsilon_0}{\pi\sqrt{4x^2-g^2}} & \text{if $x>g/2$}.
	\end{cases}
\end{equation}
The surface potential~\eqref{eq:gappot} matches the desired electrode potentials on the electrodes ($|x|>g/2$), and the surface charge density~\eqref{eq:gapscd} vanishes in the gap ($|x|<g/2$) as required. In fact Eq.~\eqref{eq:gappot} gives the \emph{only} potential which satisfies both of these conditions, and it coincides with the gap potential as determined by numerical finite-element calculations. However, in what follows we will see that this potential is not sufficient to describe arbitrary gaps.

As Eq.~\eqref{eq:gappot} gives the surface potential in the \emph{entire} $x y$ plane, we can now use Eq.~\eqref{eq:Phi23} to calculate its associated three-dimensional electric potential
\begin{multline}
	\label{eq:gapff}
	\Phi_g\se{gap}(\vect{r}) = \frac12
	+ \frac{1}{\pi}\arg\left(\sqrt{1+\mathcal{Z}^2}+\mathcal{Z}\right)\\
	= \frac12 + \frac{\vartheta}{\pi} -\frac{g^2 \sin (2 \vartheta)}{16 \pi  r^2} + \mathcal{O}[(g/r)^4],
\end{multline}
with $x=r\sin\vartheta$, $|z|=r\cos\vartheta$, and $\mathcal{Z}=\frac{|z|+i x}{g/2}=\frac{r e^{i\vartheta}}{g/2}$. The first two terms in the series expansion are due to the infinite electrodes, and the remainder to the finite gap.

In the same geometry the surface electric potential
\begin{equation}
	\label{eq:ellpot}
	\phi_g\se{pol}(x) = \begin{cases}
		\sqrt{1-\left(\frac{x}{g/2}\right)^2} & \text{if $|x|<g/2$}\\
		0 & \text{if $|x|\ge g/2$}
	\end{cases}
\end{equation}
leads to the surface charge density
\begin{equation}
	\label{eq:ellscd}
	\sigma_g\se{pol}(x) = \begin{cases}
		\frac{4 \varepsilon_0}{g} & \text{if $|x|<g/2$}\\
		\frac{4 \varepsilon_0}{g}\left( 1-\frac{2|x|}{\sqrt{4x^2-g^2}} \right) & \text{if $|x|>g/2$}.
	\end{cases}
\end{equation}
While the isolated straight gap of Fig.~\ref{fig:infgap} is fully described by Eq.~\eqref{eq:gappot}, the ``polarization potential'' of Eq.~\eqref{eq:ellpot} will appear in more complicated electrode setups. It vanishes on the electrodes ($|x|>g/2$), and the associated charge density is constant inside the gap ($|x|<g/2$). Whenever an unphysical charge density is induced within a gap, usually by nearby electrodes via Eq.~\eqref{eq:phi2sigma}, we can add a surface potential inspired by Eq.~\eqref{eq:ellpot} in order to bring the in-gap charge density back to zero without modifying the potential on the electrodes. The associated three-dimensional electric potential is
\begin{equation}
	\label{eq:polff}
	\Phi_g\se{pol}(\vect{r}) = \Re\left(\sqrt{1+\mathcal{Z}^2}-\mathcal{Z}\right)
	= \frac{g \cos \vartheta}{4r} + \mathcal{O}[(g/r)^3],
\end{equation}
where $\Re$ refers to the real part. Comparing the far-field expansions in Eq.~\eqref{eq:gapff} and Eq.~\eqref{eq:polff} we already see that whenever a gap polarization potential similar to Eq.~\eqref{eq:ellpot} is present its far field will dominate over that of the gap interpolation potential~\eqref{eq:gappot} because of its slower decay with $r$.

The two gap potentials~\eqref{eq:gappot} and~\eqref{eq:ellpot} will be shown to suffice for describing thin gaps by comparison to numerical calculations (see Fig.~\ref{fig:susceptibility}). We will therefore not need any further potentials in the following sections.

\subsection{two parallel straight gaps}
\label{sec:strip}

\begin{figure}
	\includegraphics[width=8.5cm]{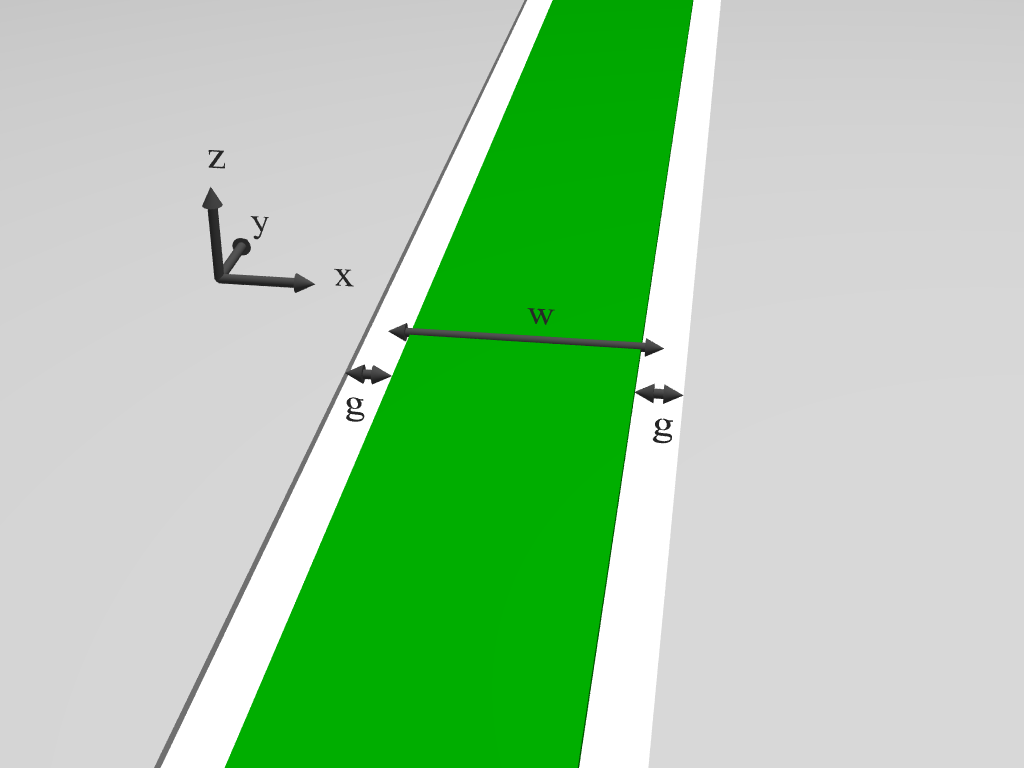}
	\caption{Perspective view of an infinitely long strip electrode of width $w$ at unit potential (green) within an infinite grounded plane (gray), separated by gaps of width $g$. The strip width $w$ is measured between gap centers.}
	\label{fig:infstrip}
\end{figure}

As an application of the surface potentials~\eqref{eq:gappot} and~\eqref{eq:ellpot} we study a strip electrode of width $w$ at unit potential separated from an infinite grounded plane by two gaps of width $g\ll w$ (Fig.~\ref{fig:infstrip}). The surface potential is approximately
\begin{multline}
	\label{eq:strippot}
	\phi_{w,g}\se{strip}(x) \approx \phi_g\se{gap}(x+\frac{w}{2}) - \phi_g\se{gap}(x-\frac{w}{2})\\
	+ \alpha\times\left[
	\phi_g\se{pol}(x+\frac{w}{2}) + \phi_g\se{pol}(x-\frac{w}{2})
	\right],
\end{multline}
where $\alpha$ is a parameter to be determined. It is not \emph{a priori} clear that $\alpha=0$ as was the case for the isolated gap in Section~\ref{sec:infinitegap}: for electrodes differing from the setup of Fig.~\ref{fig:infgap} the potential at the center of the gap need not necessarily be the average of the two adjoining electrode potentials. Electrodes which are farther away, as well as the curvature of the gap (see Section~\ref{sec:ringgaps}), can induce such a polarization potential since there are no free charges within the gap to counteract these influences. We determine the correct value for $\alpha$ by setting the surface charge densities at the centers of the two gaps to zero, as required physically:
\begin{equation}
	\label{eq:stripsigma}
	\sigma_{w,g}\se{strip}(x=\pm\frac{w}{2}) = \frac{4\varepsilon_0}{\pi} \left[
	\frac{1-2\pi\alpha w/g}{\sqrt{4w^2-g^2}}
	+\frac{2\pi\alpha}{g}
	\right]=0,
\end{equation}
giving
\begin{equation}
	\label{eq:susceptibility}
	\alpha = \frac{g}{2\pi(w-\sqrt{4w^2-g^2})} = -\frac{g}{2\pi w} + \mathcal{O}[(g/w)^3].
\end{equation}
In Fig.~\ref{fig:susceptibility} we will show that this potential matches that determined by numerical simulations for $g\ll w$.

The gap contribution to the electric potential far from the strip electrode is dominated by the gap polarization component,
\begin{multline}
	\label{eq:stripff}
	\Phi_{w,g}\se{strip}(r,\vartheta) = \left( 1 + \frac{\pi\alpha g}{2w} \right) \frac{w \cos\vartheta}{\pi r} + \mathcal{O}[(w/r)^3]\\
	\approx \left( 1 - \frac{g^2}{4w^2} \right) \frac{w \cos\vartheta}{\pi r}.
\end{multline}
This setup demonstrates that when taking gaps between electrodes into account, it is crucial to calculate the gap polarization since its contribution to the electric potential is generally more important than that of the gap interpolation potential~\eqref{eq:gappot}. Furthermore the gap-width dependent prefactor of $1-g^2/(4w^2)$ in Eq.~\eqref{eq:stripff} gives a first quantitative idea of how important gap potentials are. Even for $g=w/2$, where the width of the gaps is equal to the width of the strip electrode, its value is still 94\%.

We have limited this discussion to nulling the charge density at the \emph{center} of the gap. For increased accuracy one can derive generalizations of the polarization potential with in-gap charge distributions proportional to successively higher powers of $x$ which can serve to null successively higher moments of the in-gap charge. However, the electric potentials associated with such higher-order gap potentials drop off faster with distance from the gap than Eq.~\eqref{eq:polff} and will therefore be neglected in this work.

\subsection{ring ion trap with gaps}
\label{sec:ringgaps}

\begin{figure}
	\includegraphics[width=8.5cm]{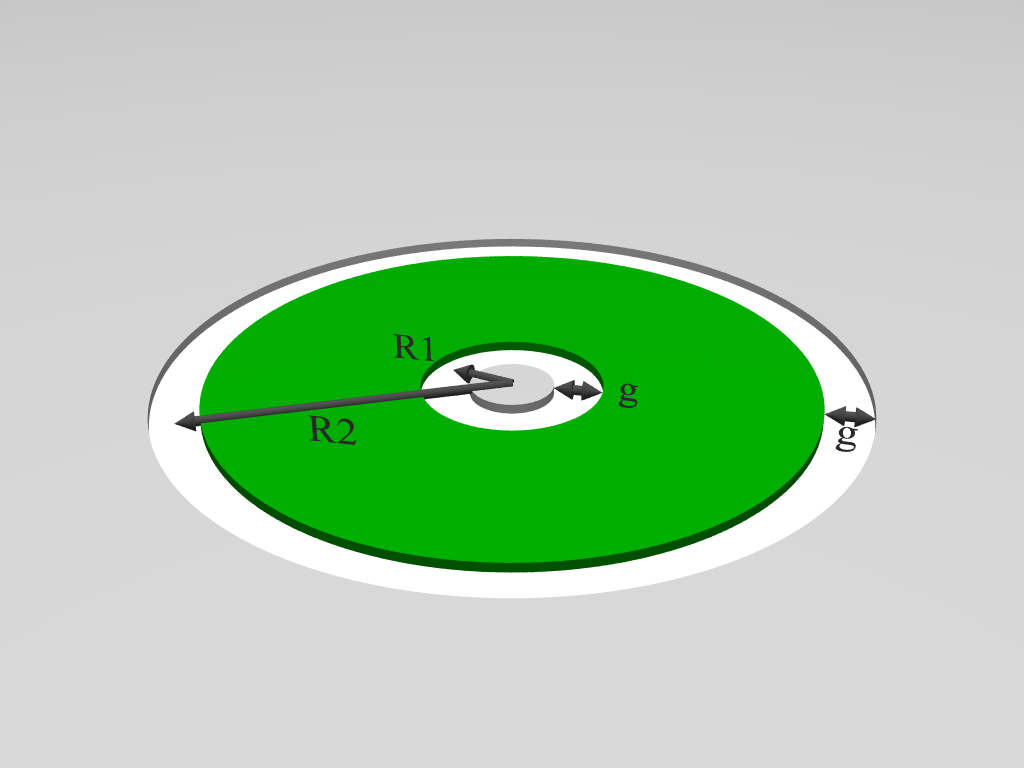}
	\caption{Perspective view of a planar ring ion (Paul) trap with gaps. An rf electrode (green) is embedded in an infinite grounded plane (gray) and surrounded by gaps of width $g$. The radii $R_{1,2}$ are measured to the gap centers.}
	\label{fig:ringgap}
\end{figure}

As a more practical application of gap potentials we study an optimized radio-frequency ring ion trap~\cite{Wesenberg2008,Schmied2009} in the presence of gaps between ground and rf electrodes (Fig.~\ref{fig:ringgap}). Assuming thin gaps and inspired by the results of Section~\ref{sec:strip} we take the cylindrically symmetric scaled in-plane potential as
\begin{widetext}
\begin{equation}
	\label{eq:ringpot}
	\phi\se{ring}(r) = \begin{cases}
		0 & \text{if $r\le R_1-\frac{g}{2}$}\\
		\phi_g\se{gap}(r-R_1) + \alpha_1 \phi_g\se{pol}(r-R_1) & \text{if $|r-R_1|<\frac{g}{2}$}\\
		1 & \text{if $R_1+\frac{g}{2} \le r \le R_2-\frac{g}{2}$}\\
		\phi_g\se{gap}(R_2-r) + \alpha_2 \phi_g\se{pol}(r-R_2) & \text{if $|r-R_2|<\frac{g}{2}$}\\
		0 & \text{if $r\ge R_2+\frac{g}{2}$}.
	\end{cases}
\end{equation}
\end{widetext}
The surface charge densities at the gap centers are, to leading orders in the gap width $g$,
\begin{subequations}
\label{eq:ringsigma}
\begin{align}
	\sigma\se{ring}(R_1) & \approx -2\varepsilon_0\left[
	-\frac{K\left(\frac{4 R_1R_2}{(R_1+R_2)^2}\right)}{\pi(R_1+R_2)}
	-\frac{E\left(\frac{4 R_1R_2}{(R_1+R_2)^2}\right)}{\pi(R_2-R_1)}
	\right.\nonumber\\
	& \left.
	-\frac{\log \left(\frac{g}{32 R_1}\right)}{2 \pi  R_1}
	-\frac{2\alpha_1}{g}
	+\frac{g R_2 \alpha_2 E\left(-\frac{4 R_1 R_2}{(R_2-R_1)^2}\right)}{2(R_1+R_2)^2 (R_2-R_1)}
   \right]\\
	\sigma\se{ring}(R_2) & \approx -2\varepsilon_0\left[
	\frac{K\left(\frac{4 R_1R_2}{(R_1+R_2)^2}\right)}{\pi(R_1+R_2)}
	-\frac{E\left(\frac{4 R_1R_2}{(R_1+R_2)^2}\right)}{\pi(R_2-R_1)}
	\right.\nonumber\\
	& \left.
	+\frac{\log \left(\frac{g}{32 R_2}\right)}{2 \pi  R_2}
	-\frac{2\alpha_2}{g}
	+\frac{g R_1 \alpha_1 E\left(-\frac{4 R_1 R_2}{(R_2-R_1)^2}\right)}{2(R_1+R_2)^2 (R_2-R_1)}
   \right],
\end{align}
\end{subequations}
where $K(m)$ and $E(m)$ are complete elliptic integrals of the first and second kind, respectively. Setting $\sigma\se{ring}(R_1)=\sigma\se{ring}(R_2)=0$ allows us to compute the amplitudes $\alpha_1$ and $\alpha_2$ necessary for nulling the surface charge densities at the centers of the two gaps, to leading orders in $g$.

\begin{figure}
	\includegraphics[width=8.5cm]{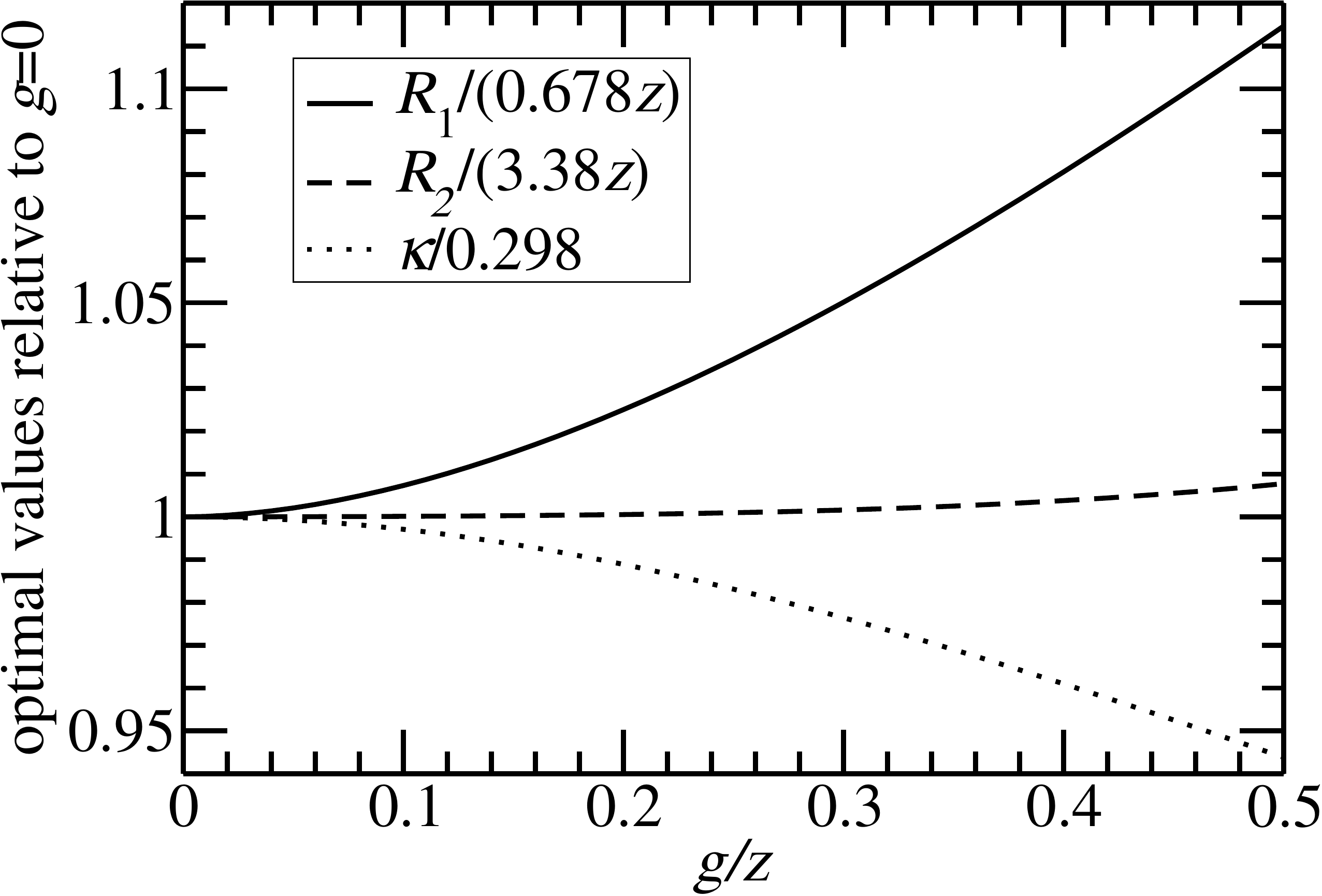}
	\caption{Optimized planar ring trap with gapped electrodes, trapping at height $z$ above the electrode plane. As a function of gap width $g$, the optimal radii and the maximal achievable curvature $\kappa$ differ from the idealized result at $g=0$.}
	\label{fig:optimalring}
\end{figure}

We do not have a closed expression for the 3D potential due to Eq.~\eqref{eq:ringpot}; but on the $z$ axis ($r=0$) it takes the simple form (to leading orders in $g$)
\begin{multline}
	\label{eq:ringpotaxis}
	\Phi\se{ring}(r=0,z) = \left[ \frac{|z|}{\sqrt{R_1^2+z^2}}-\frac{|z|}{\sqrt{R_2^2+z^2}}\right]\\
	+ \frac{\pi g |z|}{4} \left[
	\frac{\alpha_1 R_1}{(R_1^2+z^2)^{3/2}} + \frac{\alpha_2 R_2}{(R_2^2+z^2)^{3/2}} \right]\\
	+ \frac{g^2 |z|}{16} \left[
	\frac{2R_1^2-z^2}{(R_1^2+z^2)^{5/2}} - \frac{2R_2^2-z^2}{(R_2^2+z^2)^{5/2}} \right]+\ldots
\end{multline}
Similar to the system in Section~\ref{sec:strip}, for small gap widths $g$ the influence of the gap polarization potentials [second term in Eq.~\eqref{eq:ringpotaxis}] is found to dominate over that of the gap interpolation potential [third term in Eq.~\eqref{eq:ringpotaxis}].

Eq.~\eqref{eq:ringpotaxis} allows us to calculate the optimal rf ring radii as a function of gap width. The ponderomotive pseudopotential of an ion of mass $m$ and charge $q$ moving in an electric field $\Phi(\vect{r},t) = U\si{rf}\cos(\Omega\si{rf}t)\times\Phi\se{scaled}(\vect{r})$ is~\cite{Leibfried2003}
\begin{equation}
	\Psi(\vect{r}) = \frac{q^2 U\si{rf}^2 \|\vect{\nabla}\Phi\se{scaled}(\vect{r})\|^2}{4m\Omega\si{rf}^2}.
\end{equation}
For a given gap width $g$ and desired trapping height $z$ we determine the radii $R_{1,2}$ for which $\Psi(r=0,z)=0$ and which maximize the dimensionless trap curvature~\cite{Schmied2009}
\begin{equation}
	\label{eq:kappa}
	\kappa = z^2\left| \det H \Phi\se{scaled}(\vect{r}) \right|\si{trap}^{1/3}
\end{equation}
defined through the Hessian determinant of the scaled potential at the trapping site. Fig.~\ref{fig:optimalring} shows these quantities relative to their well-known values at the gapless limit $g=0$. We find that for finite gap widths $g$ the inner radius $R_1$ must be slightly increased; but the limiting value $R_1(g=0)=0.678z$ always lies \emph{within} the gap. The same is true for the outer radius, whose optimized value is even less sensitive to the presence of a gap. Finally, the trap curvature $\kappa$ is bound to decrease from its gap-free value, but this decrease is limited to a few percent for reasonable gap widths.

\subsection{arbitrary gap shapes}
\label{sec:arbitrary}

For arbitrary gap shapes, as they naturally appear \eg\ in surface electrode optimizations~\cite{Schmied2009}, simple homogeneous parametrizations such as Eqs.~\eqref{eq:strippot} and~\eqref{eq:ringpot} are not applicable. Instead, as long as the gaps are sufficiently smooth and have radii of curvature much larger than the gap width, we propose the following scheme. The center of each gap is parametrized by a curve $\vect{\gamma}(t)$ winding counterclockwise around an rf electrode, and a normalized direction $\vect{e}(t)=\left[ \begin{smallmatrix} 0&-1\\1&0 \end{smallmatrix} \right]\cdot\vect{\gamma}'(t)/\|\vect{\gamma}'(t)\|$ is defined at each point perpendicular to the gap direction. Each point $\vect{p}$ within the gap can now be described by local coordinates: $\vect{p}(t,u)=\vect{\gamma}(t)+u\vect{e}(t)$. Inspired by the results of the preceding sections we propose to parametrize the scaled gap potential as
\begin{equation}
	\label{eq:arbitrarygappot}
	\phi\se{gap}(t,u) = \phi_{g(t)}\se{gap}(u) + \alpha(t)\times\phi_{g(t)}\se{pol}(u)
\end{equation}
whenever $|u|\le g(t)/2$. As long as $\alpha(t)$ is a sufficiently smooth function of $t$, with variations on length scales much larger than the gap width $g(t)$, this potential will be a good representation of the true gap potential as determined by numerical inversion techniques. The full in-plane potential is given by a term like Eq.~\eqref{eq:arbitrarygappot} for each gap curve plus the potential on the electrodes. The exact way in which gap intersections and forks are incorporated into the above parametrization is largely irrelevant, since they represent zero-dimensional objects in the electrode plane and their 3D electrostatic potentials thus decrease faster with distance than those of the corresponding one-dimensional gap potentials, Eqs.~\eqref{eq:gapff} and~\eqref{eq:polff}.

As can be seen from comparing Eqs.~\eqref{eq:stripsigma} and~\eqref{eq:ringsigma}, the surface charge density at the gap center depends strongly on the gap shape and curvature. For general setups the surface charge density at the gap centers must be calculated using an approximation such as Eq.~\eqref{eq:phi2sigma_approx}, giving a functional dependence of the $\sigma\se{center}(t)$ on the functions $\alpha(t)$. A numerical inversion then yields the necessary $\alpha(t)$ for nulling the surface charge densities at all gap centers. Since this inversion procedure involves only one-dimensional gaps it is expected to be much faster than a full BEM calculation.

\subsection{electrode thickness and substrate permittivity}
\label{sec:thickness}

\begin{figure}
	\includegraphics[width=8.5cm]{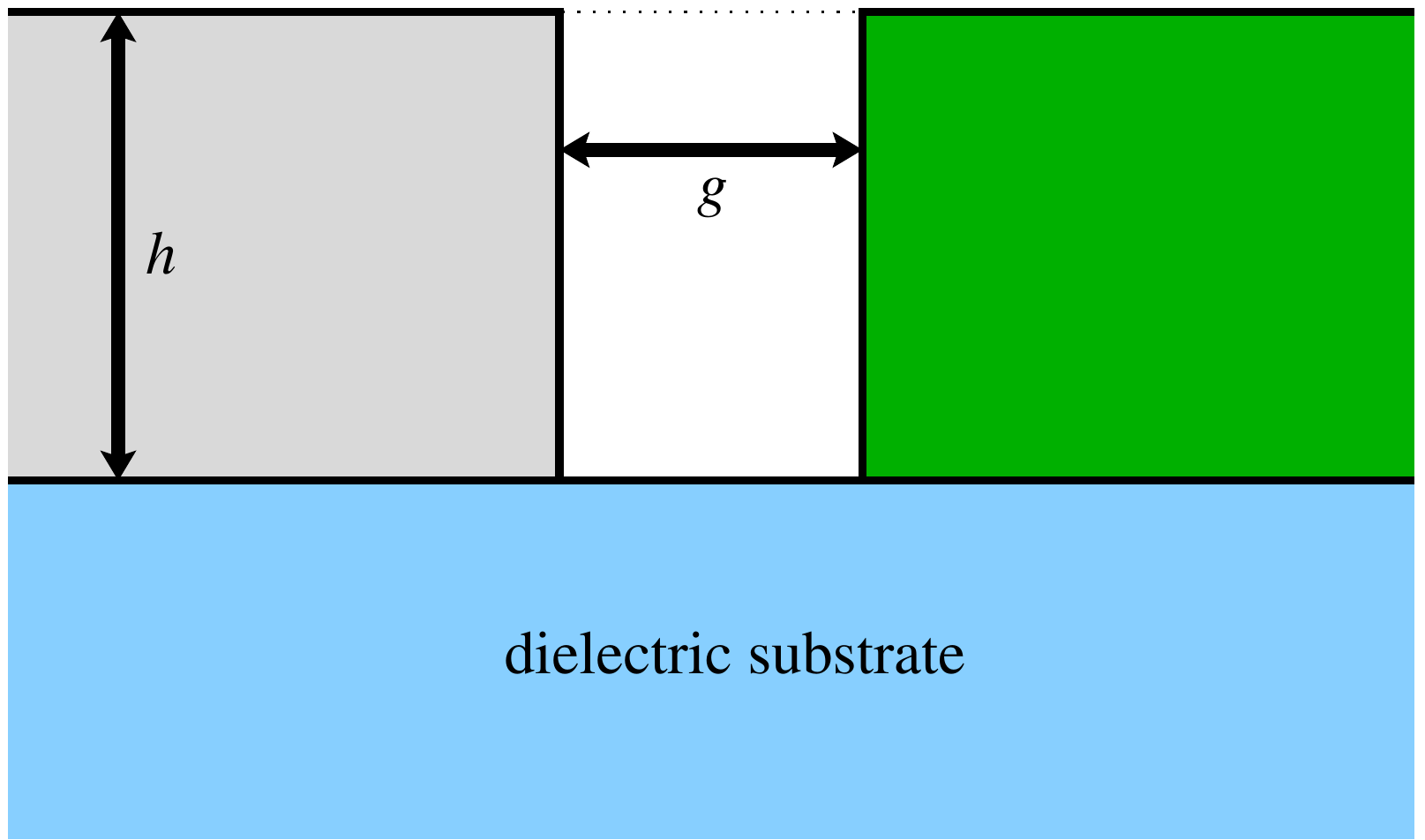}
	\caption{Cross section through a gap of width $g$ between two coplanar electrodes of thickness $h$ mounted on a substrate of dielectric constant $\varepsilon\si{s}$. The gap susceptibilities of Fig.~\ref{fig:susceptibility} are computed in the top plane indicated by a dotted line.}
	\label{fig:finitegap}
\end{figure}

One of the main assumptions of this section is that the surface electrodes are infinitely thin and lie in one plane. This assumption allowed us to define a surface charge density $\sigma(x,y)$ in the electrode plane, which could then be set to zero outside of the metallic electrodes in order to determine the surface potentials. In realistic setups, however, electrodes have a finite thickness and are often mounted on a dielectric substrate. Under these circumstances numerical inversion techniques are necessary to determine exact electric potentials. But an approximate method can nonetheless be derived in the spirit of the preceding sections. In Eqs.~\eqref{eq:stripff} and~\eqref{eq:ringpotaxis} we had found that the far-field of the electrode gaps is dominated by the ``gap polarization'', whose amplitude $\alpha$ must in general be determined by an inversion procedure (Section~\ref{sec:arbitrary}). Here we suggest that to leading order the cross sectional shape of the gap (Fig.~\ref{fig:finitegap}) merely introduces an effective \emph{gap susceptibility} which scales the gap polarization potential by a numerical prefactor.

\begin{figure}
	\includegraphics[width=8.5cm]{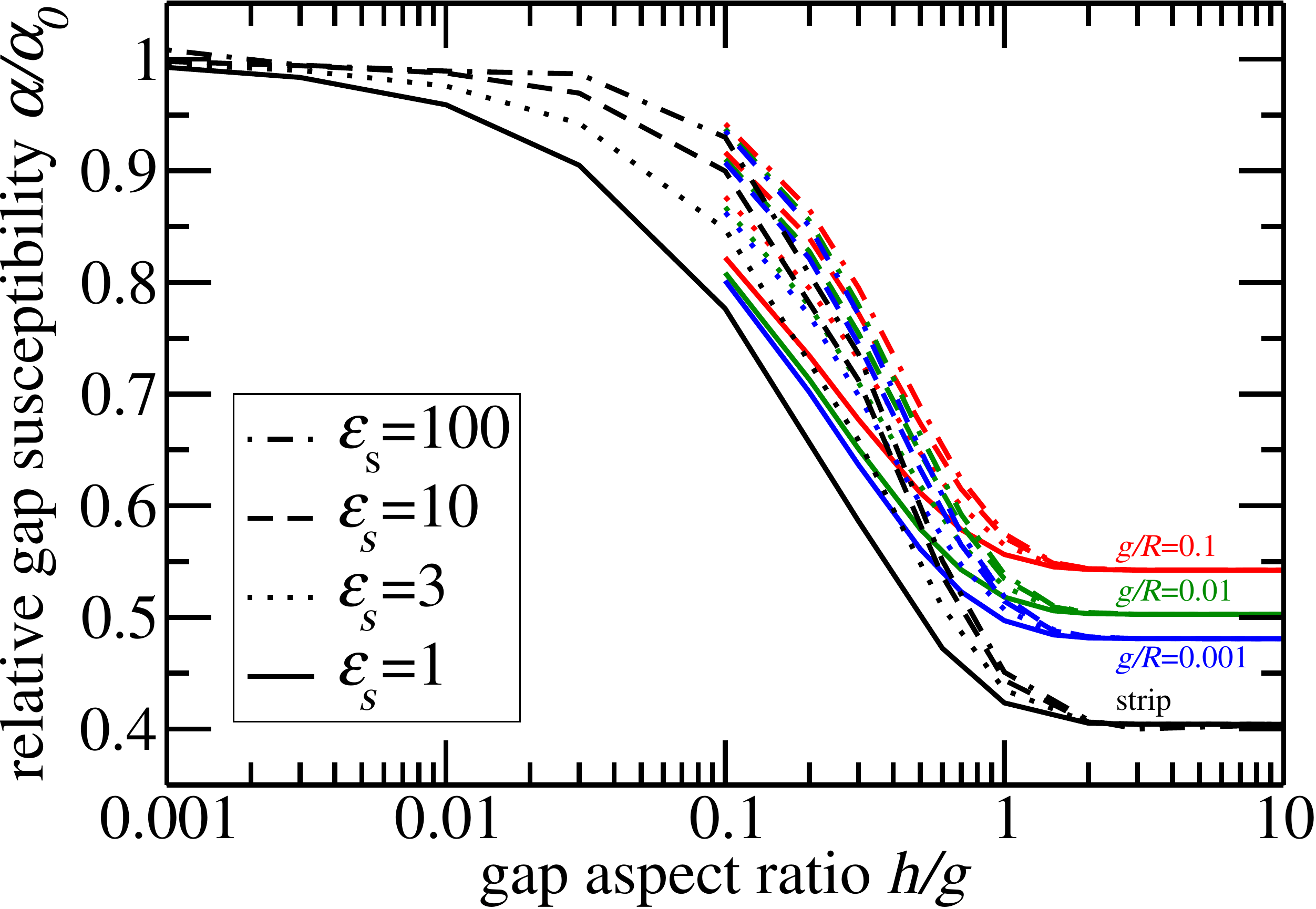}
	\caption{Relative gap susceptibility as a function of electrode aspect ratio (horizontal axis: ratio of electrode thickness $h$ to gap width $g$), substrate dielectric constant $\varepsilon\si{s}$ (line styles), and gap curvature (colors). Black: translation-invariant numerical calculations of the setup of Fig.~\ref{fig:infstrip} with thick electrodes as in Fig.~\ref{fig:finitegap}; $\alpha_0=-g/(2\pi w)$ [Eq.~\eqref{eq:susceptibility}]. For $g\ll w$, $\alpha/\alpha_0$ does not depend on the strip width $w$. Colors: cylindrically symmetric numerical calculations of the setup of Fig.~\ref{fig:ringgap} with $R_1=0$ and vanishing inner gap; $\alpha_0=\frac{g^2}{16R^2}\ln\frac{g}{32R}$.}
	\label{fig:susceptibility}
\end{figure}

We have used two finite-element computer codes (translation invariant and cylindrically symmetric) to calculate the effects of finite gap aspect ratios and substrate dielectric constants (Fig.~\ref{fig:finitegap}) on the surface electric potential (along the dotted line in Fig.~\ref{fig:finitegap}). As in thin-electrode calculations, this surface potential in the gap can be decomposed into a gap interpolation component similar to Eq.~\eqref{eq:gappot} and a polarization potential equal to Eq.~\eqref{eq:ellpot}, whose amplitude $\alpha$ is found from a line integral of the numerically determined electric potential across the gap. Comparing these numerical values of $\alpha$ to those known from analytic thin-electrode calculations ($\alpha_0$) yields the relative gap susceptibility $\alpha/\alpha_0$. In complex setups this relative gap susceptibility cannot be expected to depend only on local gap properties. But since Eq.~\eqref{eq:phi2sigma} is dominated by a local quasi-singular contribution, we expect the leading-order dependences of $\alpha/\alpha_0$ to be local, namely (i) the gap aspect ratio $h/g$, (ii) the substrate dielectric constant $\varepsilon\si{s}$, and (iii) the gap curvature. Fig.~\ref{fig:susceptibility} summarizes our numerical results showing these three dependences for several highly symmetric systems with uniform gaps. The main dependence is found to be on the gap aspect ratio. In the limit of large gap aspect ratios ($h/g\gtrsim 0.7$) all of our results are in the vicinity of $\alpha/\alpha_0\approx0.5$, which leads us to propose the following simplified model for gap simulations with significant aspect ratios: after numerically determining the gap polarizations $\alpha_0$ by the thin-electrode inversion method of Section~\ref{sec:arbitrary}, we simply multiply them by 0.5 before evaluating their electric potential. While this method does not take the exact gap curvature into account and incorrectly attributes equal weights to the influences of adjoining and far-off electrodes on the gap polarization through Eq.~\eqref{eq:phi2sigma}, it nonetheless gives a good approximation of the gap potential. Keeping in mind that the effects of gaps on three-dimensional potentials are generally small (see Fig.~\ref{fig:optimalring}), such an approximation is expected to be sufficient for the design phase of surface electrodes.

\section{finite electrode plane}
\label{sec:infinite}

The second approximation usually made for surface-electrode ion traps~\cite{Wesenberg2008,House2008,Schmied2009} is that of an infinite grounded electrode plane surrounding the trapping region. But in realistic systems, the electrode plane is necessarily finite. In this section we attempt to take this fact into account for a particular electrode geometry in order to estimate its importance.

Instead of assuming an infinite grounded plane we start from a planar grounded circular electrode $\Theta$ of radius $S$ surrounded by vacuum. All rf and dc electrodes will be placed on $\Theta$. The reason why we choose $\Theta$ to be circular is to simplify the following analysis in cylindrical coordinates; but substituting a rectangular $\Theta$ for a circular one of similar size is not expected to change the conclusions of this section qualitatively, and the tools developed in this section can thus be applied to a larger set of problems. For a given in-plane potential $\phi(r,\vartheta)$ on $\Theta$, determined by the shapes and potentials of the electrodes placed on $\Theta$, the 3D potential $\Phi(r,\vartheta,z)$ must satisfy the following conditions:
\begin{enumerate}
	\item[(i)] The surface limit $\lim_{z\to0} \Phi(r,\vartheta,z) = \phi(r,\vartheta)$ must be valid for all $r\le S$. However, for $r>S$ we have no restrictions on the surface potential, as by assumption there is no electrode to fix the potential. The in-plane potential at $r>S$ is to be determined.
	\item[(ii)] The surface charge density $\sigma(r,\vartheta) = -2\varepsilon_0 \lim_{z\to 0^+} \partial \Phi(r,\vartheta,z)/\partial z$ must vanish for $r>S$ since by assumption there is no electrode to carry the charge density. But for $r<S$ it is to be determined.
\end{enumerate}
As in Section~\ref{sec:infinitegap} there is a unique potential which satisfies both of these constraints. Here we derive this potential for an infinitesimally small ``pixel'' of nonzero in-plane potential on $\Theta$ at radius $\rho$, given by $\phi(r,\vartheta) = \rho^{-1} \delta(r-\rho)\delta(\vartheta)$ with $0< \rho< S$. The surface charge density is calculated from Eq.~\eqref{eq:phi2sigma}. For $r>\rho$ it can be multipole-expanded as~\cite{footnote1}
\begin{multline}
	\label{eq:multipole}
	\sigma(r>\rho,\vartheta) = -\frac{\varepsilon_0}{\pi(\rho^2+r^2-2\rho r\cos\vartheta)^{3/2}}\\
	= -\sum_{(m,n)}
	\frac{2^{3-\delta_{m,0}} \varepsilon_0 (\frac{n-m-1}{2})_{\frac12} (\frac{n+m-1}{2})_{\frac12}\rho^{n-3}}{\pi^2}
	\frac{\cos(m\vartheta)}{r^n},
\end{multline}
where the sum is over $m=0,1,2,\ldots$ and $n=m+3,m+5,m+7,\ldots$; $(a)_n=\Gamma(a+n)/\Gamma(a)$ is the Pochhammer symbol. This surface charge density must be brought to zero for $r>S$ in order to satisfy condition (ii), but without modifying the surface potential for $r<S$. We notice that surface charge densities of the form
\begin{multline}
	\label{eq:surfacechargecompensation}
	\hat{\sigma}_{S,m,n}(r,\vartheta) = \frac{\cos(m\vartheta)}{r^n}\times\\
	\begin{cases}
		-\frac{(\frac{n+m+2}{2})_{-\frac32}}{2\sqrt{\pi}}
		\frac{r^{n+m}}{S^{n+m}} \twoFone(\frac32,\frac{n+m}{2}; \frac{n+m+2}{2}; \frac{r^2}{S^2}) & \text{if $r<S$}\\
		1 & \text{if $r\ge S$}
	\end{cases}
\end{multline}
yield in-plane potentials through Eq.~\eqref{eq:sigma2phi} of
\begin{multline}
	\hat{\phi}_{S,m,n}(r,\vartheta) = -\cos(m\vartheta)\times \frac{2\sqrt{\pi}(\frac{n+m}{2})_{-\frac12}}{S^n(n-m-1)}\times\\
	\begin{cases}
		0 & \text{if $r<S$}\\
		\frac{\sqrt{r^2-S^2}}{(r/S)^m}
		\Re \twoFone(1,\frac{n-m}{2};\frac{n-m+1}{2};\frac{r^2}{S^2}) & \text{if $r\ge S$}
	\end{cases}
\end{multline}
where $\twoFone$ is Gauss' hypergeometric function and $\Re$ refers to its real part. Adding such multipole charge densities to Eq.~\eqref{eq:multipole} thus allows us to null the charge density for $r\ge S$ while leaving the associated surface potential on the electrode $\Theta$ ($r<S$) unchanged. While we do not have a closed expression for the three-dimensional electric potentials due to the charge densities of Eq.~\eqref{eq:surfacechargecompensation}, their series expansions for $r^2+z^2<S^2$ are
\begin{multline}
	\label{eq:multipoleseriespot}
	\hat{\Phi}_{S,m,n}(r,\vartheta,z) = -\cos(m\vartheta) \sum_{j,k=0}^{\infty}
	\frac{(-4)^{k+1}}{S^n} \left( \frac{r}{S} \right)^{m+2j}\\ \left( \frac{|z|}{S} \right)^{1+2k}
	\frac{ (j+1)_{k+\frac12} (j+m+1)_k(\frac{n+m}{2})_{-\frac12}}{[n+m+2(j+k)](1+2k)!}.
\end{multline}
Combining Eqs.~\eqref{eq:multipole} and~\eqref{eq:multipoleseriespot} gives the three-dimensional potential of the in-plane potential ``pixel'' at radius $\rho$ on the finite electrode $\Theta$,
\begin{multline}
	\label{eq:surrpot}
	\tilde{G}_{S,\rho}(r,\vartheta,z) = \frac{|z|}{2\pi(\rho^2+r^2+z^2-2\rho r \cos\vartheta)^{3/2}}\\
	+ \sum_{i,j,k,m=0}^{\infty}
	(-1)^k 2^{3+2k-\delta_{m,0}}\\
	\times
	\frac{
	(i+1)_{\frac12}
	(j+1)_{k+\frac12}
	(j+m+1)_k
	}{\pi^3[3+2(i+j+k+m)] (1+2k)!}\\
	\times \frac{1}{S^2} \left( \frac{\rho}{S} \right)^{m+2i} \left( \frac{r}{S} \right)^{m+2j} \left( \frac{|z|}{S} \right)^{1+2k} \cos(m\vartheta).
\end{multline}
We recognize the first term in Eq.~\eqref{eq:surrpot} as the Green's function of Eq.~\eqref{eq:Green} in cylindrical coordinates, to which the 3D potential must converge in the limit $S\to\infty$. The second term is the correction due to the finiteness of the electrode $\Theta$. The quadruple sum over Pochhammer symbols in Eq.~\eqref{eq:surrpot} can be reduced to a triple sum over hypergeometric functions by performing one of the summations analytically; however we have found that this does not make numerical evaluation faster in practice. Arbitrary electrode patterns within $\Theta$ can be constructed from Eq.~\eqref{eq:Phi23} by interpreting Eq.~\eqref{eq:surrpot} as a modified Green's function which takes the finite electrode $\Theta$ into account. The radius of convergence of Eq.~\eqref{eq:surrpot} is $r^2+z^2<S^2$, which is sufficient for practical simulations. For large $S$ the Green's function in Cartesian coordinates is approximately
\begin{equation}
	\tilde{G}_{S}(x,y,z) =
	\frac{|z|}{2\pi(x^2+y^2+z^2)^{3/2}}
	+ \frac{|z|}{3\pi^2 S^3} + \mathcal{O}(S^{-4}),
\end{equation}
adding to Eq.~\eqref{eq:Green} a small constant out-of-plane electric field which can be easily incorporated into numerical calculations.

\subsection{ring ion trap on finite electrode}
\label{sec:finitering}

\begin{figure}
	\includegraphics[width=8.5cm]{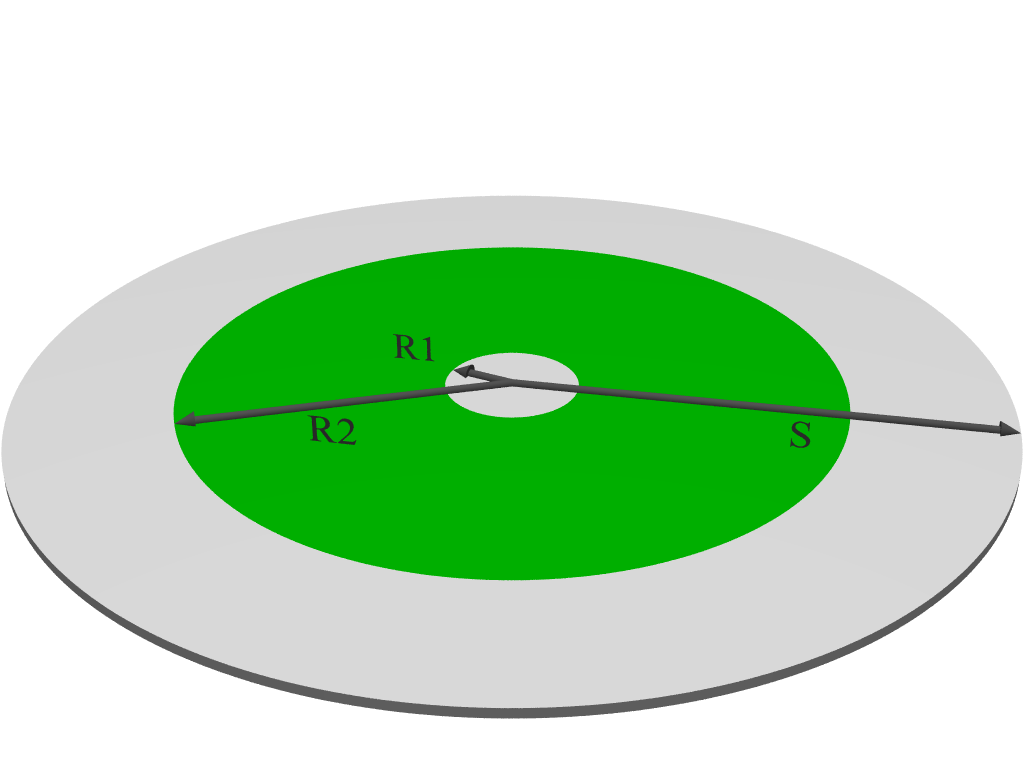}
	\caption{Perspective view of a planar ring ion (Paul) trap on a finite electrode. An rf electrode (green) is embedded in a finite grounded plane (gray) and surrounded by vacuum.}
	\label{fig:ringsurr}
\end{figure}

\begin{figure}
	\includegraphics[width=8.5cm]{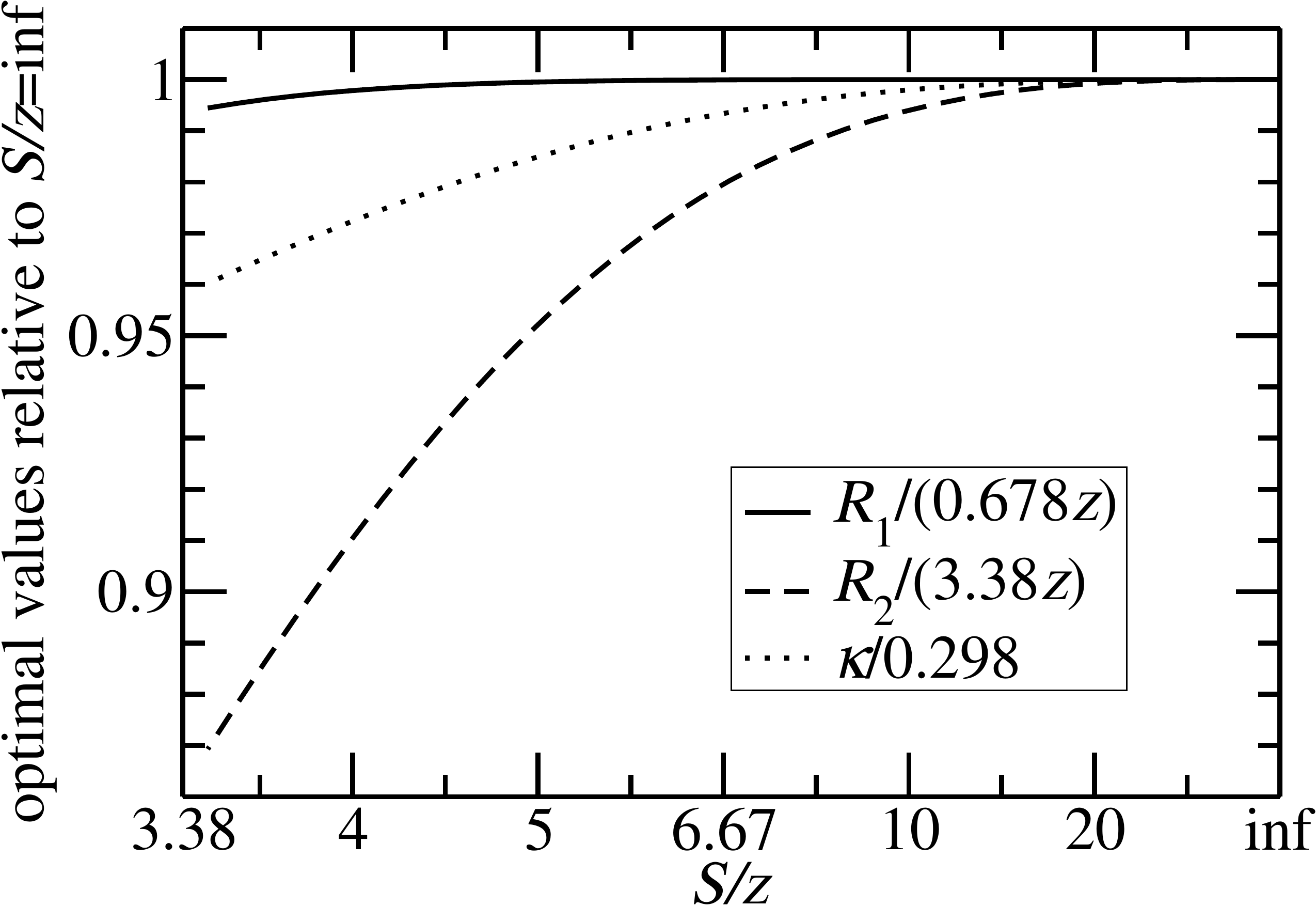}
	\caption{Optimized planar ring trap with gapped electrodes, trapping at height $z$ above the electrode plane. As a function of the reciprocal of the grounded electrode radius $S$, the optimal radii and the maximal achievable curvature $\kappa$ differ from the idealized result at $S=\infty$.}
	\label{fig:ringsurropt}
\end{figure}

As an example of a surface electrode pattern surrounded by a finite grounded electrode, we again study the optimized ring ion trap. Fig.~\ref{fig:ringsurr} shows the schematic setup consisting of a ring-shaped rf electrode on a grounded disc of finite radius. Its three-dimensional electric potential
\begin{equation}
	\Phi_S(r,z) = \int_{R_1}^{R_2} \rho \text{d}\rho \int_0^{2\pi}\text{d}\vartheta \tilde{G}_{S,\rho}(r,\vartheta,z)
\end{equation}
can be simplified on the axis ($r=0$) to
\begin{multline}
	\label{eq:onaxisringS}
	\Phi_S(r=0,z) = \frac{|z|}{\sqrt{R_1^2+z^2}}-\frac{|z|}{\sqrt{R_2^2+z^2}}\\
	+ \sum_{i,k=0}^{\infty} \frac{(-1)^k \left[\left(\frac{R_2}{S}\right)^{2+2i}-\left(\frac{R_1}{S}\right)^{2+2i}\right]}
	{(\frac32+i+k)\pi^{3/2} (i+\frac32)_{\frac12}}
	\times \left(\frac{|z|}{S}\right)^{1+2k}.
\end{multline}
As in Section~\ref{sec:ringgaps}, Eq.~\eqref{eq:onaxisringS} allows us to calculate the optimal rf ring radii as a function of the size of $\Theta$: for given radius $S$ and desired trapping height $z$ we determine the radii $R_{1,2}$ which maximize the dimensionless trap curvature $\kappa$ defined in Eq.~\eqref{eq:kappa}. Fig.~\ref{fig:ringsurropt} shows these quantities relative to their well-known values in the limit of an infinite electrode plane $S=\infty$. While the inner radius $R_1$ is very insensitive to the finiteness of the substrate, the outer radius shows a rather strong dependence. This can be understood by the physical proximity of the outer radius to the substrate edge. As in Fig.~\ref{fig:optimalring}, $\kappa$ is reduced from its ideal value ($S\to\infty$) by only a few percent.

\section{conclusions}

After deriving general frameworks we have estimated the influence of finite gaps (Section~\ref{sec:ringgaps}) and a finite surrounding electrode (Section~\ref{sec:finitering}) on the optimization of a ring-shaped Paul ion trap constructed from surface electrodes. These influences were found to be small in realistic situations: even extreme examples, such as a gap width of half the trapping height and a vanishing grounded electrode surrounding the rf trapping electrode, each give reductions in the dimensionless trap curvature of the order of only 5\%. A similar small reduction was found in the far-field of a strip electrode (Section~\ref{sec:strip}). More realistic gap geometries including a finite gap aspect ratio and a substrate dielectric (Fig.~\ref{fig:finitegap}) further reduce the influence of gap potentials by a factor of two.

We conclude that for many practical calculations of surface electrode designs it is a safe assumption to neglect the influence of both gaps and surroundings, since fabrication tolerances and stray fields will often introduce uncertainties in excess of these effects; practical implementations of surface ion traps necessitate control electrodes for correcting these uncertainties, and can simultaneously correct for design approximations such as those stemming from the effects quantified in this work.

For those situations where higher accuracy is required, however, we have presented a method for calculating the effects of finite gaps and finite surrounding electrodes. This method provides fast estimates of the leading-order influences of gaps and finite electrodes on three-dimensional electric potentials generated by surface electrodes. The small size of these influences suggest that the two effects can likely be estimated separately, since the cross-polarizations of gaps and the ``outside'' plane are expected to be on the order of a few percent of the dominant polarization effects. After the design phase of a surface electrode setup using this approximate method, a full BEM calculation can then serve to validate the approximations and provide a fully accurate system description.

\section{acknowledgments}

This work was supported by the European Union through the SCALA integrated project, and by the German Research Foundation (DFG) through the cluster of excellence \emph{Munich Center for Advanced Photonics} (MAP) and through program FOR 635.


\begin{thebibliography}{17}
\expandafter\ifx\csname natexlab\endcsname\relax\def\natexlab#1{#1}\fi
\expandafter\ifx\csname bibnamefont\endcsname\relax
  \def\bibnamefont#1{#1}\fi
\expandafter\ifx\csname bibfnamefont\endcsname\relax
  \def\bibfnamefont#1{#1}\fi
\expandafter\ifx\csname citenamefont\endcsname\relax
  \def\citenamefont#1{#1}\fi
\expandafter\ifx\csname url\endcsname\relax
  \def\url#1{\texttt{#1}}\fi
\expandafter\ifx\csname urlprefix\endcsname\relax\def\urlprefix{URL }\fi
\providecommand{\bibinfo}[2]{#2}
\providecommand{\eprint}[2][]{\url{#2}}

\bibitem[{\citenamefont{Cirac and Zoller}(1995)}]{Cirac1995}
\bibinfo{author}{\bibfnamefont{J.~I.} \bibnamefont{Cirac}} \bibnamefont{and}
  \bibinfo{author}{\bibfnamefont{P.}~\bibnamefont{Zoller}},
  \bibinfo{journal}{Phys. Rev. Lett.} \textbf{\bibinfo{volume}{74}},
  \bibinfo{pages}{4091} (\bibinfo{year}{1995}).

\bibitem[{\citenamefont{Schmidt-Kaler et~al.}(2003)\citenamefont{Schmidt-Kaler,
  H{\"a}ffner, Riebe, Gulde, Lancaster, Deuschle, Becher, Roos, Eschner, and
  Blatt}}]{SchmidtKaler2003}
\bibinfo{author}{\bibfnamefont{F.}~\bibnamefont{Schmidt-Kaler}},
  \bibinfo{author}{\bibfnamefont{H.}~\bibnamefont{H{\"a}ffner}},
  \bibinfo{author}{\bibfnamefont{M.}~\bibnamefont{Riebe}},
  \bibinfo{author}{\bibfnamefont{S.}~\bibnamefont{Gulde}},
  \bibinfo{author}{\bibfnamefont{G.~P.~T.} \bibnamefont{Lancaster}},
  \bibinfo{author}{\bibfnamefont{T.}~\bibnamefont{Deuschle}},
  \bibinfo{author}{\bibfnamefont{C.}~\bibnamefont{Becher}},
  \bibinfo{author}{\bibfnamefont{C.~F.} \bibnamefont{Roos}},
  \bibinfo{author}{\bibfnamefont{J.}~\bibnamefont{Eschner}}, \bibnamefont{and}
  \bibinfo{author}{\bibfnamefont{R.}~\bibnamefont{Blatt}},
  \bibinfo{journal}{Nature} \textbf{\bibinfo{volume}{422}},
  \bibinfo{pages}{408} (\bibinfo{year}{2003}).

\bibitem[{\citenamefont{Blatt and Wineland}(2008)}]{Blatt2008}
\bibinfo{author}{\bibfnamefont{R.}~\bibnamefont{Blatt}} \bibnamefont{and}
  \bibinfo{author}{\bibfnamefont{D.}~\bibnamefont{Wineland}},
  \bibinfo{journal}{Nature} \textbf{\bibinfo{volume}{453}},
  \bibinfo{pages}{1008} (\bibinfo{year}{2008}).

\bibitem[{\citenamefont{Buluta and Nori}(2009)}]{Buluta2009}
\bibinfo{author}{\bibfnamefont{I.}~\bibnamefont{Buluta}} \bibnamefont{and}
  \bibinfo{author}{\bibfnamefont{F.}~\bibnamefont{Nori}},
  \bibinfo{journal}{Science} \textbf{\bibinfo{volume}{326}},
  \bibinfo{pages}{108} (\bibinfo{year}{2009}).

\bibitem[{\citenamefont{Dehmelt}(1990)}]{Dehmelt1990}
\bibinfo{author}{\bibfnamefont{H.}~\bibnamefont{Dehmelt}},
  \bibinfo{journal}{Rev. Mod. Phys.} \textbf{\bibinfo{volume}{62}},
  \bibinfo{pages}{525} (\bibinfo{year}{1990});
\bibinfo{author}{\bibfnamefont{W.}~\bibnamefont{Paul}}, \bibinfo{journal}{Rev.
  Mod. Phys.} \textbf{\bibinfo{volume}{62}}, \bibinfo{pages}{531}
  (\bibinfo{year}{1990}).

\bibitem[{\citenamefont{Leibfried et~al.}(2003)\citenamefont{Leibfried, Blatt,
  Monroe, and Wineland}}]{Leibfried2003}
\bibinfo{author}{\bibfnamefont{D.}~\bibnamefont{Leibfried}},
  \bibinfo{author}{\bibfnamefont{R.}~\bibnamefont{Blatt}},
  \bibinfo{author}{\bibfnamefont{C.}~\bibnamefont{Monroe}}, \bibnamefont{and}
  \bibinfo{author}{\bibfnamefont{D.}~\bibnamefont{Wineland}},
  \bibinfo{journal}{Rev. Mod. Phys.} \textbf{\bibinfo{volume}{75}},
  \bibinfo{pages}{281} (\bibinfo{year}{2003}).

\bibitem[{\citenamefont{H{\"a}ffner et~al.}(2005)\citenamefont{H{\"a}ffner,
  H{\"a}nsel, Roos, Benhelm, Chek{-}al{-}kar, Chwalla, K{\"o}rber, Rapol,
  Riebe, Schmidt et~al.}}]{Haeffner2005}
\bibinfo{author}{\bibfnamefont{H.}~\bibnamefont{H{\"a}ffner}},
  \bibinfo{author}{\bibfnamefont{W.}~\bibnamefont{H{\"a}nsel}},
  \bibinfo{author}{\bibfnamefont{C.~F.} \bibnamefont{Roos}},
  \bibinfo{author}{\bibfnamefont{J.}~\bibnamefont{Benhelm}},
  \bibinfo{author}{\bibfnamefont{D.}~\bibnamefont{Chek{-}al{-}kar}},
  \bibinfo{author}{\bibfnamefont{M.}~\bibnamefont{Chwalla}},
  \bibinfo{author}{\bibfnamefont{T.}~\bibnamefont{K{\"o}rber}},
  \bibinfo{author}{\bibfnamefont{U.~D.} \bibnamefont{Rapol}},
  \bibinfo{author}{\bibfnamefont{M.}~\bibnamefont{Riebe}},
  \bibinfo{author}{\bibfnamefont{P.~O.} \bibnamefont{Schmidt}},
  \bibnamefont{et~al.}, \bibinfo{journal}{Nature}
  \textbf{\bibinfo{volume}{438}}, \bibinfo{pages}{643} (\bibinfo{year}{2005}).

\bibitem[{\citenamefont{Seidelin et~al.}(2006)\citenamefont{Seidelin,
  Chiaverini, Reichle, Bollinger, Leibfried, Britton, Wesenberg, Blakestad,
  Epstein, Hume et~al.}}]{Seidelin2006}
\bibinfo{author}{\bibfnamefont{S.}~\bibnamefont{Seidelin}},
  \bibinfo{author}{\bibfnamefont{J.}~\bibnamefont{Chiaverini}},
  \bibinfo{author}{\bibfnamefont{R.}~\bibnamefont{Reichle}},
  \bibinfo{author}{\bibfnamefont{J.~J.} \bibnamefont{Bollinger}},
  \bibinfo{author}{\bibfnamefont{D.}~\bibnamefont{Leibfried}},
  \bibinfo{author}{\bibfnamefont{J.}~\bibnamefont{Britton}},
  \bibinfo{author}{\bibfnamefont{J.~H.} \bibnamefont{Wesenberg}},
  \bibinfo{author}{\bibfnamefont{R.~B.} \bibnamefont{Blakestad}},
  \bibinfo{author}{\bibfnamefont{R.~J.} \bibnamefont{Epstein}},
  \bibinfo{author}{\bibfnamefont{D.~B.} \bibnamefont{Hume}},
  \bibnamefont{et~al.}, \bibinfo{journal}{Phys. Rev. Lett.}
  \textbf{\bibinfo{volume}{96}}, \bibinfo{pages}{253003}
  (\bibinfo{year}{2006}).

\bibitem[{\citenamefont{Amini et~al.}(2008)\citenamefont{Amini, Britton,
  Leibfried, and Wineland}}]{Amini2008}
\bibinfo{author}{\bibfnamefont{J.~M.} \bibnamefont{Amini}},
  \bibinfo{author}{\bibfnamefont{J.}~\bibnamefont{Britton}},
  \bibinfo{author}{\bibfnamefont{D.}~\bibnamefont{Leibfried}},
  \bibnamefont{and} \bibinfo{author}{\bibfnamefont{D.~J.}
  \bibnamefont{Wineland}}, \bibinfo{journal}{arXiv:0812.3907v1}
  (\bibinfo{year}{2008}).

\bibitem[{\citenamefont{Wesenberg}(2008)}]{Wesenberg2008}
\bibinfo{author}{\bibfnamefont{J.~H.} \bibnamefont{Wesenberg}},
  \bibinfo{journal}{Phys. Rev. A} \textbf{\bibinfo{volume}{78}},
  \bibinfo{pages}{063410} (\bibinfo{year}{2008}).

\bibitem[{\citenamefont{House}(2008)}]{House2008}
\bibinfo{author}{\bibfnamefont{M.~G.} \bibnamefont{House}},
  \bibinfo{journal}{Phys. Rev. A} \textbf{\bibinfo{volume}{78}},
  \bibinfo{pages}{033402} (\bibinfo{year}{2008}).

\bibitem[{\citenamefont{Schmied et~al.}(2009)\citenamefont{Schmied, Wesenberg,
  and Leibfried}}]{Schmied2009}
\bibinfo{author}{\bibfnamefont{R.}~\bibnamefont{Schmied}},
  \bibinfo{author}{\bibfnamefont{J.~H.} \bibnamefont{Wesenberg}},
  \bibnamefont{and}
  \bibinfo{author}{\bibfnamefont{D.}~\bibnamefont{Leibfried}},
  \bibinfo{journal}{Phys. Rev. Lett.} \textbf{\bibinfo{volume}{102}},
  \bibinfo{pages}{233002} (\bibinfo{year}{2009}).

\bibitem[{\citenamefont{Wrobel}(2002)}]{Wrobel}
\bibinfo{author}{\bibfnamefont{L.~C.} \bibnamefont{Wrobel}},
  \emph{\bibinfo{title}{The Boundary Element Method}}, vol.~\bibinfo{volume}{1}
  (\bibinfo{publisher}{John Wiley \& Sons}, \bibinfo{year}{2002}).

\bibitem[{\citenamefont{Liu and Nishimura}(2006)}]{Liu2006}
\bibinfo{author}{\bibfnamefont{Y.~J.} \bibnamefont{Liu}} \bibnamefont{and}
  \bibinfo{author}{\bibfnamefont{N.}~\bibnamefont{Nishimura}},
  \bibinfo{journal}{Engineering Analysis with Boundary Elements}
  \textbf{\bibinfo{volume}{30}}, \bibinfo{pages}{371} (\bibinfo{year}{2006}).

\bibitem[{\citenamefont{Oliveira and Miranda}(2001)}]{Oliveira2001}
\bibinfo{author}{\bibfnamefont{M.~H.} \bibnamefont{Oliveira}} \bibnamefont{and}
  \bibinfo{author}{\bibfnamefont{J.~A.} \bibnamefont{Miranda}},
  \bibinfo{journal}{Eur. J. Phys.} \textbf{\bibinfo{volume}{22}},
  \bibinfo{pages}{31} (\bibinfo{year}{2001}).

\bibitem[{\citenamefont{Jackson}(1999)}]{Jackson}
\bibinfo{author}{\bibfnamefont{J.~D.} \bibnamefont{Jackson}},
  \emph{\bibinfo{title}{Classical Electrodynamics}} (\bibinfo{publisher}{John
  Wiley \& Sons}, \bibinfo{year}{1999}), \bibinfo{edition}{3rd} ed.

\bibitem{footnote1}
\bibinfo{note}{At $(r,\vartheta)=(\rho,0)$ there is a singularity in $\sigma$, which we are not interested in here.}

\end{thebibliography}
\end{document}